\documentclass[nofootinbib,aps,11pt]{revtex4-1}
\usepackage{graphicx}
\usepackage{hyperref}
\usepackage{cancel}
\usepackage{amssymb}
\usepackage{textcomp}
\usepackage{amsmath}
\usepackage{bm}
\usepackage{times}
\usepackage{epsfig}
\usepackage{color}
\usepackage{ulem}
\newcommand{\GeV}{{\rm GeV}}

\newcommand{\TeV}{{\rm TeV}}

\newcommand{\fb}{{\rm fb}}
\newcommand{\pb}{{\rm pb}}

\newcommand{\SM}{{\rm SM}}
\newcommand{\BR}{{\rm BR}}

\begin{document}
\title{\LARGE Interpretation of 750 GeV diphoton excess at LHC in singlet extension of color-octet neutrino mass model}
\bigskip
\author{Ran Ding~$^{1}$}
\email{dingran@mail.nankai.edu.cn}
\author{Zhi-Long Han~$^{2}$}
\email{hanzhilong@mail.nankai.edu.cn}
\author{Yi Liao~$^{3,2,1}$}
\email{liaoy@nankai.edu.cn}
\author{Xiao-Dong Ma~$^{2}$}
\email{maxid@mail.nankai.edu.cn}
\affiliation{
$^1$ Center for High Energy Physics, Peking University, Beijing 100871, China
\\
$^2$~School of Physics, Nankai University, Tianjin 300071, China
\\
$^3$ State Key Laboratory of Theoretical Physics, Institute of Theoretical Physics,
Chinese Academy of Sciences, Beijing 100190, China}
\date{\today}

\begin{abstract}

We propose that the possible $750~\GeV$ diphoton excess can be explained in the color-octet neutrino mass model extended with a scalar singlet $\Phi$. The model generally contains $N_s$ species of color-octet, electroweak doublet scalars $S$ and $N_f$ species of color-octet, electroweak triplet $\chi$ or singlet $\rho$ fermions. While both scalars and fermions contribute to the production of $\Phi$ through gluon fusion, only the charged members induce the diphoton decay of $\Phi$. The diphoton rate can be significantly enhanced due to interference between the scalar and fermion loops. We show that the diphoton cross section can be from 3 to 10 $\fb$ for $\mathcal{O}(\TeV)$ color-octet particles while evading all current LHC limits.

\end{abstract}

\maketitle

%%%%%%%%%%%%%%%%%%%%%%%
\section{The Model}\label{model}
%%%%%%%%%%%%%%%%%%%%%%%

Recently, both ATLAS and CMS have found an excess in the diphoton channel around $750~\GeV$ with a width possibly $\lesssim45~\GeV$ in the LHC run-II data at $\sqrt{s}=13~\TeV$ \cite{ATLAS:2015dip,CMS:2015dxe}. The local significance of the excess is $3.6\sigma$ at $747~\GeV$ and $2.6\sigma$ at $760~\GeV$, and the global significance is $2.0\sigma$ and $1.2\sigma$ for ATLAS \cite{ATLAS:2015dip} and CMS \cite{CMS:2015dxe} respectively. While CMS previously observed a slight excess $\sim2\sigma$ around $750~\GeV$ at $\sqrt{s}=8~\TeV$ \cite{Khachatryan:2015qba}, ATLAS did not go beyond $600~\GeV$ in the same channel \cite{Aad:2014ioa}. If the excess persists with accumulation of more data in the near future, it will likely point to new physics beyond the standard model (SM). The excess has caused a burst of discussions on its possible origin in various phenomenological frameworks \cite{Diphoton,Buttazzo:2015txu,Falkowski:2015swt,Cao:2015twy,Zhang:2015uuo,Salvio:2015jgu}.

On the other hand, there is an established piece of evidence for new physics beyond SM, namely the neutrino mass; and dark matter whose gravitational evidence is robust may also originate from new particles and physics that are unknown to us so far. It would be desirable if the newly found excess is related to the same physics of neutrino mass or dark matter. In this work we try to connect the excess with physics that is responsible for neutrino mass. Since the cross section for the excess is large for such a heavy resonance, it seems natural that the resonance is produced and decays through interactions with some new particles that participate in strong and electromagnetic interactions. In this context the color-octet model of neutrino mass \cite{FileviezPerez:2009ud} stands out, in which the octet particles generating radiative neutrino mass would also couple to the resonance field resulting in its strong production and electromagnetic decay.

Since the nature of the resonance, named $\Phi$ below, such as parity and spin is completely unknown, we treat it in the simplest manner from the point of view of effective field theory. Assuming it is an electroweak singlet of spin zero, its relevant effective interaction with color-octet scalars is
\begin{eqnarray}\label{VPhi}
V(\Phi,H,S_x)\supset \mu_{rs}\Phi S_r^{a\dagger}S_s^a,
\label{trilinear}
\end{eqnarray}
where the octet fields $S_r^a$ have the same electroweak quantum numbers as the SM Higgs field $H$ with $a$ referring to color and $r$ enumerating $N_s\ge 1$ such scalars. We note two points in passing. First, the interaction entering radiative neutrino mass is proportional to $S_r^{a\dagger}HS_s^{a\dagger}H$ while the coupling in eq. (\ref{trilinear}) does not contribute. Second, the interaction may originally be a combination of trilinear and quartic terms when the field develops an expectation value. We are aware that a large $\mu$ could be potentially dangerous for vacuum stability, but a simple estimation based on the boundedness from below indicates that for TeV-scale octet scalars and appropriate values of scalar couplings, there is no threat on vacuum stability as long as $\mu$ does not far exceed TeV scale. For renormalization-group improved analyses on the issue, see Ref.~\cite{He:2013tla} in the presence of color octet scalars, Ref.~\cite{Cao:2015twy} with color octet scalars coupled to a singlet scalar $\Phi$, and Refs.~\cite{Zhang:2015uuo,Salvio:2015jgu} with the singlet $\Phi$ being a scalar or pseudoscalar that couples to a vector-like fermion. The octet fermions carry no hypercharge and can be a singlet, $\rho_x^a$, or a triplet $\chi_x^a$, of $SU(2)_L$, where $x$ enumerates $N_f$ species of fermions. The Yukawa coupling for radiative neutrino mass is proportional to $\bar F_{L}\tilde S^a_r\rho_x^a$ or $\bar F_L\chi_x^a\tilde S_r^a$, where $F_L$ is the SM left-handed lepton doublet, $\tilde S=i\sigma_2S^*$, and $\chi$ is written in a $2\times 2$ matrix form. The relevant Yukawa coupling for our purpose here is simply,
\begin{eqnarray}\label{Yrho}
{\cal L}_\rho & \supset & -y_{xy}\Phi\overline{\rho_x^a}\rho_y^a,
\label{yukawa1}
\end{eqnarray}
or
\begin{eqnarray}\label{Ychi}
{\cal L}_\chi & \supset & -y_{xy}\Phi\textrm{tr}\overline{\chi_x^a}\chi_y^a.
\label{yukawa2}
\end{eqnarray}
For simplicity we will assume a diagonal and universal Yukawa coupling in numerical analysis. The neutral singlet scalar $\Phi$ in general mixes with the SM Higgs boson. While the mixing can be made small in a complete theory by arranging, for instance, a relatively small quartic coupling to the SM Higgs field, it must be phenomenologically small in order not to break the constraints established for the SM Higgs boson, see for instance Ref.~\cite{Cao:2015twy}. We will therefore neglect the mixing in our work.

%%%%%%%%%%%%%%%%%%%%%%%
\section{The Diphoton Excess and LHC Constraints}\label{diphoton}
%%%%%%%%%%%%%%%%%%%%%%%

The rich phenomenology for color-octet particles at LHC has been extensively studied in the literature \cite{coloroctet}. We mention some of it relevant to our study here. The Yukawa couplings of $S$ to quarks ($q$) must be small to avoid constraints from flavor physics and single production of neutral color-octet scalars at LHC \cite{Han:2010rf,Khachatryan:2015dcf,ATLAS:2015nsi}. Consequently, the effective $\Phi q\bar{q}$ coupling induced by the $S$ loop can be ignored. The direct search for pair production of $S$ in the $Zgb\bar{b}$ (with $g$ denoting gluon) final state by CMS has excluded  $m_S<625~\GeV$ at 95\% C.L. \cite{Khachatryan:2015jea}, while the CMS  search for four jets \cite{Khachatryan:2014lpa} and ATLAS search for four tops \cite{Aad:2015gdg} have excluded $m_S\lesssim830~\GeV$ at 95\% C.L.. Concerning color-octet fermions, the pair production of gluinos has been searched for at LHC in the context of simplified supersymmetric models, with their masses excluded up to $1.1-1.2~\TeV$ at 8 TeV LHC~\cite{Chatrchyan:2014lfa,Aad:2014pda}. Very recently, the analysis of 13 TeV LHC data has significantly extended the lower mass bound up to $1.6-1.8~\TeV$~\cite{ATLAS:2015glu,Khachatryan:2016kdk}. Considering these direct search constraints, we will illustrate our numerical result by assuming $m_{\chi,\rho}=2~\TeV$ and $m_S=1,~1.5~\TeV$ throughout this paper.

Currently, there exist a series of observed upper limits related to various decay final states of the singlet $\Phi$, which must be respected in our analysis. To be specific, for $m_\Phi=750~\GeV$, the searches for $jj$, $VV$, $hh$, $t\bar{t}$, and $\ell^{+}\ell^{-}$ channels at LHC $8~\TeV$ have set the bounds on the cross sections: $\sigma_{jj}<1.78~\pb$ \cite{Aad:2014aqa,CMS:2015neg}
\footnote{This upper bound for the dijet cross section corresponds to the $gg$ final state which dominates in our case. In the analysis of the bound the acceptance has been taken into account, whose number however was not available from Refs.~\cite{Aad:2014aqa,CMS:2015neg}. We will assume an acceptance of unity in our numerical analysis. A more realistic value of it will lead to a looser bound and thus permit a larger parameter space.},
$\sigma_{hh}<45~\fb$~\cite{Aad:2014yja,Aad:2015uka,Aad:2015xja},
$\sigma_{WW}<60~\fb$~\cite{Aad:2015agg}, $\sigma_{ZZ}<12~\fb$~\cite{Aad:2015kna},
$\sigma_{Z\gamma}<6~\fb$~\cite{Aad:2014fha}, $\sigma_{t\bar{t}}<300~\fb$~\cite{Aad:2015fna,Khachatryan:2015sma}, and $\sigma_{\ell^{+}\ell^{-}}<1.3~\fb$~\cite{Aad:2014cka}.
Among them, the constraints from the $hh$, $WW$, $ZZ$, and $t\bar{t}$ channels can be safely ignored due to negligible mixing between the SM Higgs $h$ and $\Phi$. As we will see later, the bounds from the $Z\gamma$ and dilepton channels can also be easily evaded.

We now consider the diphoton signature at LHC. The excess observed by ATLAS and CMS implies a theoretical diphoton rate roughly in the range $\sigma_{\gamma\gamma}\sim3-10~\fb$ depending on the narrow or wide width assumption of $\Phi$ that was obtained in Refs.~\cite{Buttazzo:2015txu,Falkowski:2015swt} by taking into account efficiency and acceptance, see for instance, the footnote 1 in Ref.~\cite{Falkowski:2015swt}. From the interactions shown in eqs. (\ref{trilinear}) and (\ref{yukawa1}) or (\ref{yukawa2}), we obtain the decay widths of $\Phi$ to the most relevant channels:
\begin{eqnarray}\label{DWrho}
\Gamma_{\Phi\rightarrow gg}^\rho & = & \frac{N_c^2\alpha_s^2m_\phi^3}{128\pi^3}\left|N_f\sum_{\rho}\frac{2y}{m_\rho}A_{1/2}(\tau_\rho)
+N_s\sum_{S^0,S^{\pm}}\frac{\mu}{m_S^2}A_0(\tau_S)\right|^2,
\\
\Gamma_{\Phi\rightarrow \gamma\gamma}^\rho & = & \frac{(N_c^2-1)^2\alpha^2m_\phi^3}{1024\pi^3}\left|N_s\sum_{S^{\pm}}\frac{\mu}{m_S^2}A_0(\tau_S)\right|^2,
\\
\Gamma_{\Phi\rightarrow Z\gamma}^\rho & = & \frac{(N_c^2-1)^2\alpha^2m_\phi^3}{512\pi^3}\left(1-\frac{m_Z^2}{m_\phi^2}\right)^3
\left(\frac{1-2s_W^2}{c_Ws_W}\right)^2\left|N_s\sum_{S^{\pm}}
\frac{\mu}{m_S^2}A_0(\tau_S^{-1},\eta_S^{-1})\right|^2,
\end{eqnarray}
for the case of singlet fermions $\rho$, and
\begin{eqnarray}\label{DWchi}
\Gamma_{\Phi\rightarrow gg}^\chi & = & \frac{N_c^2\alpha_s^2m_\phi^3}{128\pi^3}\left|N_f\sum_{\chi^0,\chi^{\pm}}\frac{2 y}{m_\chi}
A_{1/2}(\tau_\chi)+N_s\sum_{S^0,S^{\pm}}\frac{\mu}{m_S^2}A_0(\tau_S)\right|^2,\\
\Gamma_{\Phi\rightarrow \gamma\gamma}^\chi & = & \frac{(N_c^2-1)^2\alpha^2m_\phi^3}{1024\pi^3}\left|N_f\sum_{\chi^{\pm}}\frac{2 y}{m_\chi}A_{1/2}(\tau_\chi)
+N_s\sum_{S^{\pm}}\frac{\mu}{m_S^2}A_0(\tau_S)\right|^2,
\\
\nonumber
\Gamma_{\Phi\rightarrow Z\gamma}^\chi & = & \frac{(N_c^2-1)^2\alpha^2m_\phi^3}{512\pi^3}\left(1-\frac{m_Z^2}{m_\phi^2}\right)^3
\left(\frac{1-2s_W^2}{c_W s_W}\right)^2 \times
\\
& & \left|N_f\sum_{\chi^{\pm}}\frac{4 y}{m_\chi}A_{1/2}(\tau_\chi^{-1},\eta_\chi^{-1})
-N_s\sum_{S^{\pm}}\frac{\mu}{m_S^2}A_0(\tau_S^{-1},\eta_S^{-1})\right|^2,
\end{eqnarray}
for the case of triplet fermions $\chi$, where $N_c=3$ is the number of colors and $\alpha$ ($\alpha_s$) is the QED (QCD) fine structure constant. To reduce the number of parameters we will assume in numerical analysis that the $N_s$ ($N_f$) species of octet scalars (fermions) are degenerate. The loop functions $A_{0,1/2}$ of $\tau_i=m_\Phi^2/(4m_i^2)$ and of $\tau_i$ and $\eta_i=m_Z^2/(4m_i^2)$ are well known, and can be found, for instance, in Ref. \cite{Djouadi:2005gi}. Note that the decay width for the diphoton channel distinguishes between the singlet and triplet cases since the electrically neutral singlet $\rho$ does not enter. In the singlet case, the branching ratio for diphoton may receive considerable enhancement when there is a strong cancellation in the digluon channel between the octet fermions and scalars, which occurs for $y\mu<0$. In contrast, for the triplet case, the branching ratio for diphoton tends to a constant when the neutral and charged particles are almost degenerate.

In the narrow width approximation, the diphoton cross section from the $\Phi$ resonance is estimated as
\begin{equation}
\sigma_{\gamma\gamma} = \frac{\Gamma_{\Phi\to gg}}{\Gamma_{H\to gg}^\SM}
   \times \sigma^\SM(H)\times \mbox{BR}(\Phi\to\gamma\gamma),
\end{equation}
where $\Gamma_{H\to gg}^\SM$ denotes the digluon decay width of a SM-like Higgs boson $H$ scaled up to $m_H=750~\GeV$, and $\sigma^\SM(H)=735~\fb$ is its production cross section at $13~\TeV$ LHC with NNLO accuracy \cite{Heinemeyer:2013tqa,CERNReport}. For $\Gamma_{H\to gg}^\SM$, we employed the full one-loop result in Ref. \cite{Djouadi:2005gi} with a top quark mass $m_t=172~\GeV$ and neglected the bottom quark contribution. The value of $\sigma^\SM(H)$ at 13 TeV was obtained by rescaling its value ($156.8~\fb$) listed in Ref.~\cite{Heinemeyer:2013tqa} for a mass of $750~\GeV$ at 8 TeV LHC by the $gg$ parton luminosity ratio ($4.69$) from 8 TeV to 13 TeV~\cite{CERNReport}. In our scenario, the constraint in the $Z\gamma$ channel is trivially fulfilled since the decay width for $\Phi\to Z\gamma$ is always smaller than that for $\Phi\to \gamma\gamma$ in both the singlet and triplet cases. The dilepton decays of $\Phi$ are induced at one loop through the trilinear coupling in eq.~(\ref{trilinear}) and the Yukawa coupling $y^{\prime}\bar F_{L}\tilde S\rho$ or $y^{\prime}\bar F_{L}\chi\tilde S$. The same Yukawa coupling appears also in the anomalous magnetic moments and electromagnetic transitions of the charged leptons, and is thus severely constrained by current experimental bounds. Without requiring a special flavor structure, these bounds impose a universal limit $|y^{\prime}|\lesssim 0.01$. This implies that the dilepton decay widths of $\Phi$ are generally much smaller than that of diphoton. We therefore only have to consider the dijet constraint from $\Phi\rightarrow gg$. In our numerical analysis, the upper limit on $\sigma_{jj}$ at $13~\TeV$ is obtained by rescaling the limits at $8~\TeV$ through their $gg$ parton luminosity ratio, which yields $\sigma_{jj}<8.35~\pb$. We will restrict ourselves to the minimal choice that can induce two nonzero neutrino masses, namely, with the numbers of octet species being $(N_f,N_s)=(2,1)$ or $(N_f,N_s)=(1,2)$.

\begin{figure}[!htbp]
\begin{center}
\includegraphics[width=0.35\linewidth]{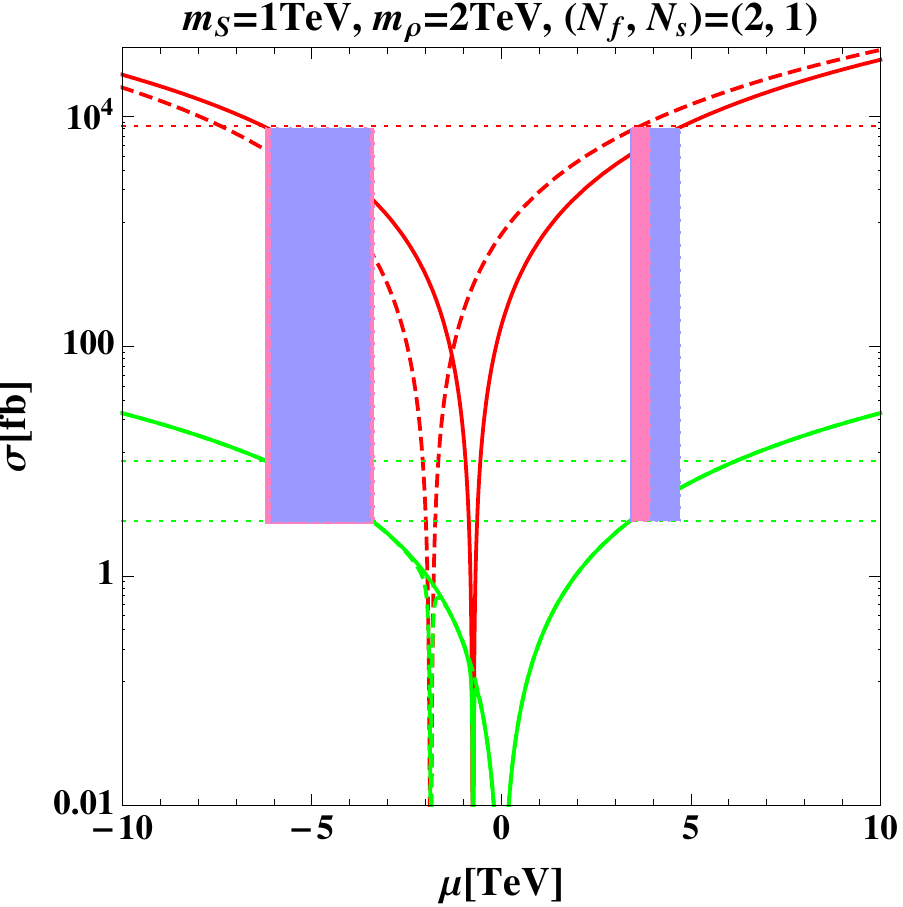}
\includegraphics[width=0.35\linewidth]{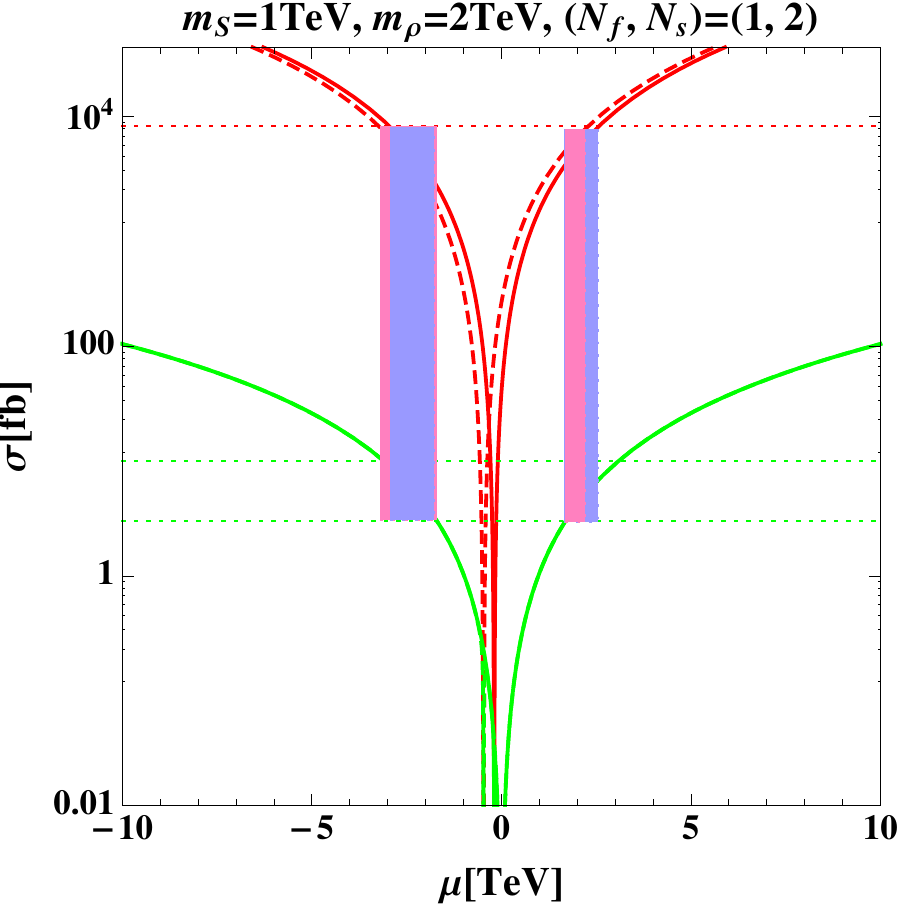}
\end{center}
\caption{Cross sections $\sigma_{gg}$ (red curves) and $\sigma_{\gamma\gamma}$ (green) are shown as a function of $\mu$ for the singlet fermion case with $m_\rho=2~\TeV$, $m_S=1~\TeV$. The left and right panels correspond to $(N_f,N_s)=(2,1),~(1,2)$ respectively, and the solid (dashed) curve assumes $y=0.2$ ($0.5$).}
\label{rhomu}
\end{figure}
\begin{figure}[!htbp]
\begin{center}
\includegraphics[width=0.35\linewidth]{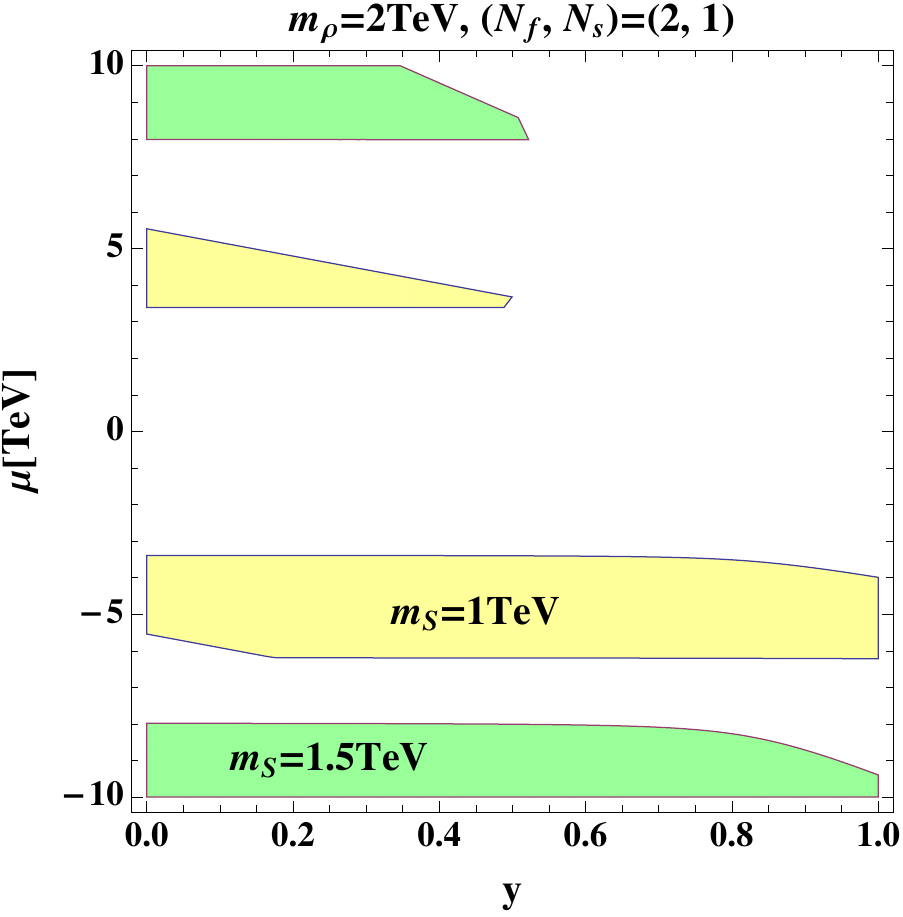}
\includegraphics[width=0.35\linewidth]{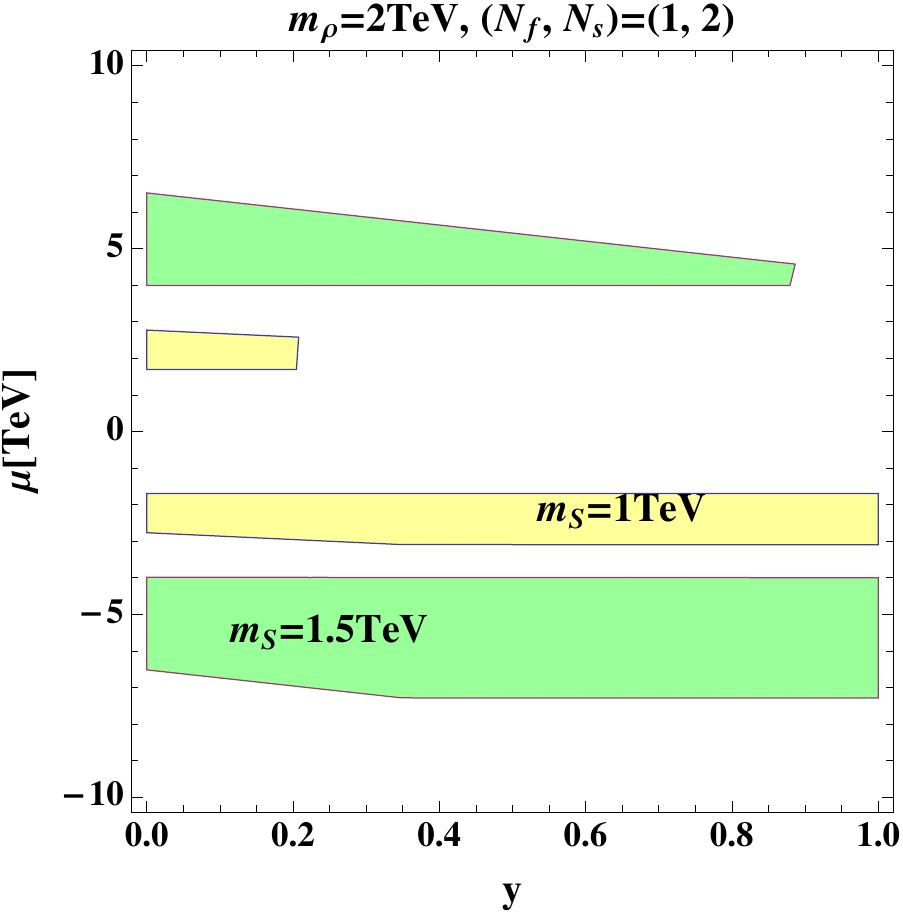}
\end{center}
\caption{Allowed parameter space in the $y-\mu$ plane for the singlet fermion case with $m_\rho=2~\TeV$ and $m_S=1~\TeV$ (yellow) or $m_S=1.5~\TeV$ (green). The left and right panels correspond to $(N_f,N_s)=(2,1),~(1,2)$ respectively.}
\label{rhoy}
\end{figure}

We first investigate the singlet fermion case. In Fig.~\ref{rhomu}, we plot the diphoton ($\sigma_{\gamma\gamma}$) and dijet ($\sigma_{gg}$) cross sections as a function of the trilinear coupling $\mu$ for the two minimal choices of neutrino mass generation. The vertical bands represent the allowed region of $\mu$ by the required diphoton excess and the upper dijet limit, with the blue (pink) band corresponding to the Yukawa coupling $y=0.2$ ($0.5$). In addition, the allowed parameter space in the $y-\mu$ plane is presented in Fig. \ref{rhoy}. From these two figures, we learn some important features:
\begin{itemize}
\item The diphoton cross section required by the excess can be reached for $\mathcal{O}(0.1)$ Yukawa coupling and $\mathcal{O}(\TeV)$ trilinear coupling $\mu$.
\item For $\mu$ positive (negative), the main constraint comes from $\sigma_{gg}$ ($\sigma_{\gamma\gamma}$). This arises from the interference effect in the $gg$ channel which is constructive for $\mu>0$ and destructive for $\mu<0$ (recalling that $y$ is always assumed to be positive), while the $\gamma\gamma$ channel has no such interference as it is contributed only by the octet scalars.
\item For a smaller Yukawa coupling $y=0.2$, $\mu$ is allowed to be either positive or negative. On the contrary, for a larger Yukawa coupling, only negative $\mu$ survives. This also results from the interference effect, and implies that the Yukawa coupling cannot be too large for a positive $\mu$.
\item As shown in Fig.~\ref{rhoy}, the allowed regions for the diphoton excess are not so sensitive to $y$, because $y$ enters in $\sigma_{\gamma\gamma}$ only indirectly through the $gg$ channel.
\item The allowed values of $|\mu|$ are smaller in the case of $(N_f,N_s)=(1,2)$ than $(N_f,N_s)=(2,1)$, simply because more charged particles contribute to the diphoton process in the former case.
\item Since the loop functions are more suppressed by heavier octet particles, a larger $|\mu|$ is demanded to fulfil the observed diphoton excess.
\end{itemize}

\begin{figure}[!htbp]
\begin{center}
\includegraphics[width=0.35\linewidth]{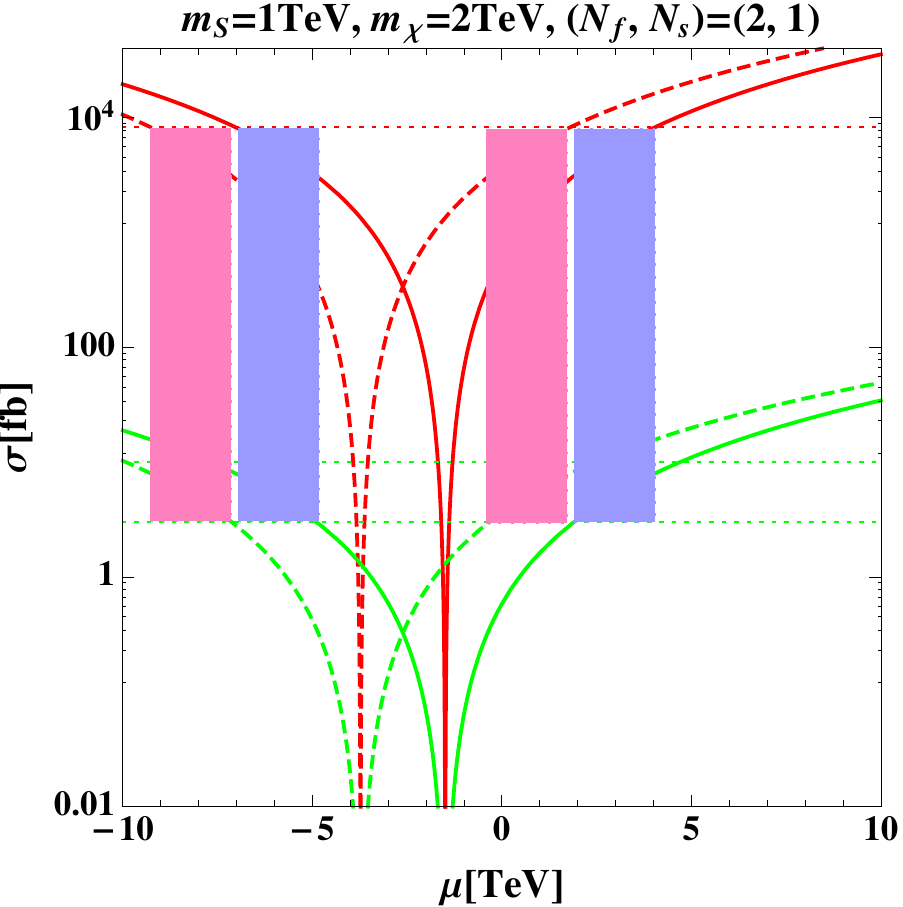}
\includegraphics[width=0.35\linewidth]{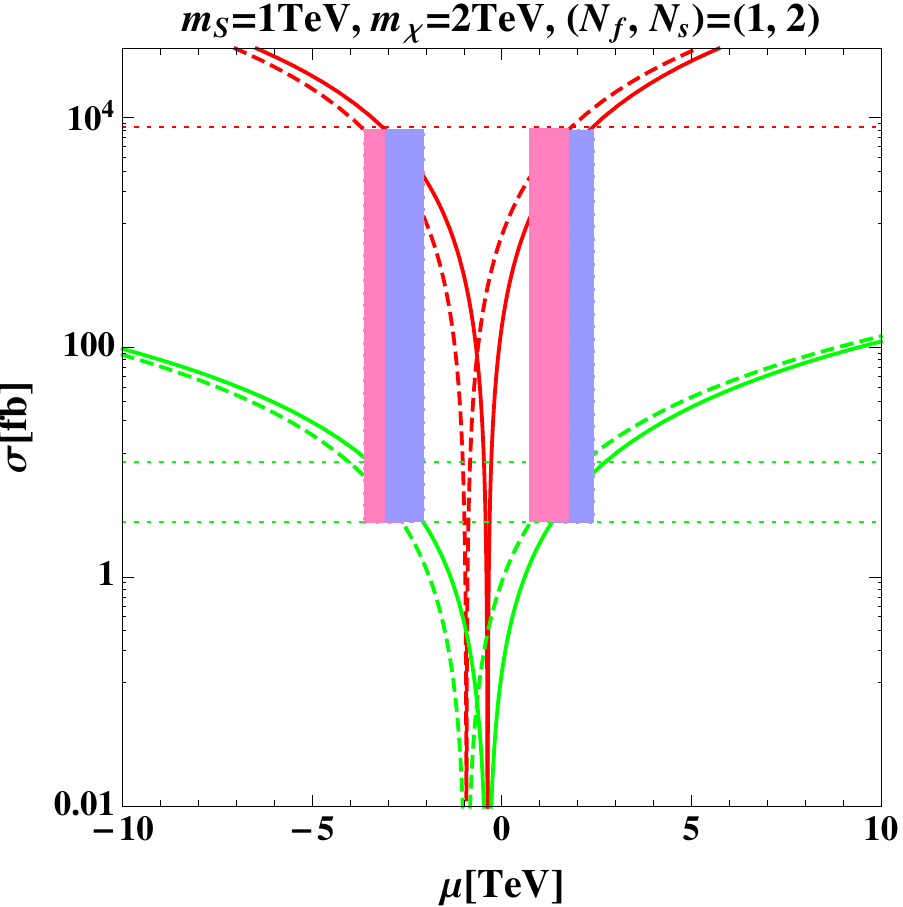}
\end{center}
\caption{Same as Fig. \ref{rhomu}, but for the triplet fermion case.}
\label{chimu}
\end{figure}

\begin{figure}[!htbp]
\begin{center}
\includegraphics[width=0.35\linewidth]{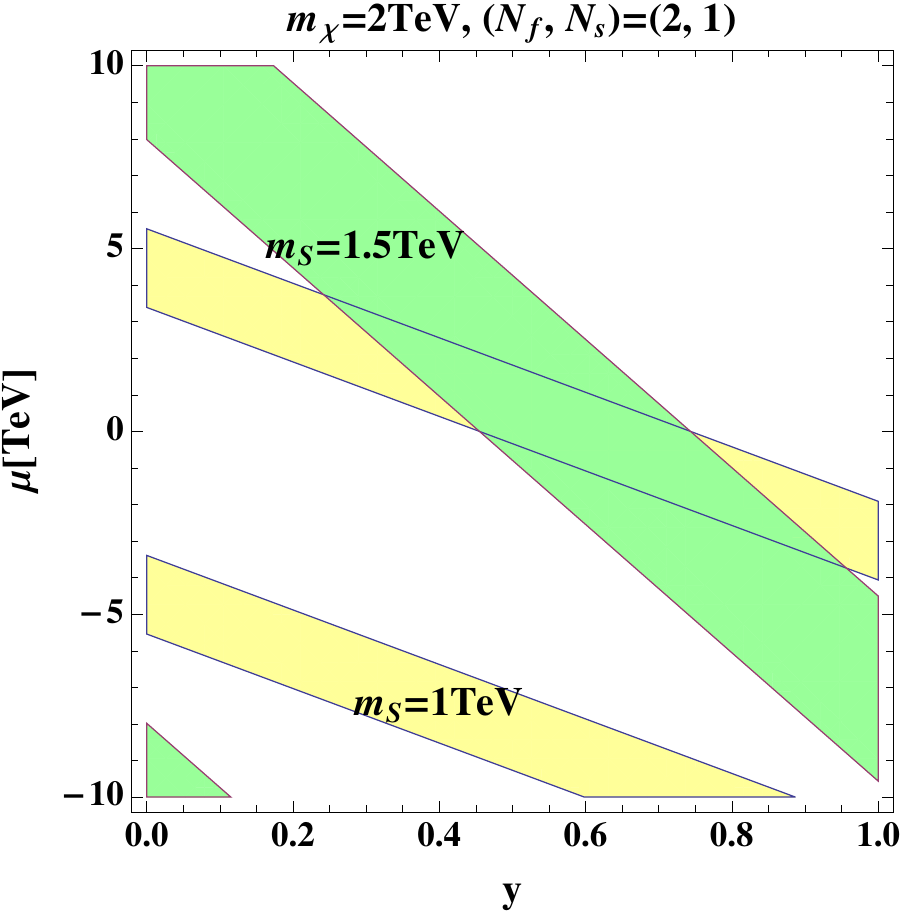}
\includegraphics[width=0.35\linewidth]{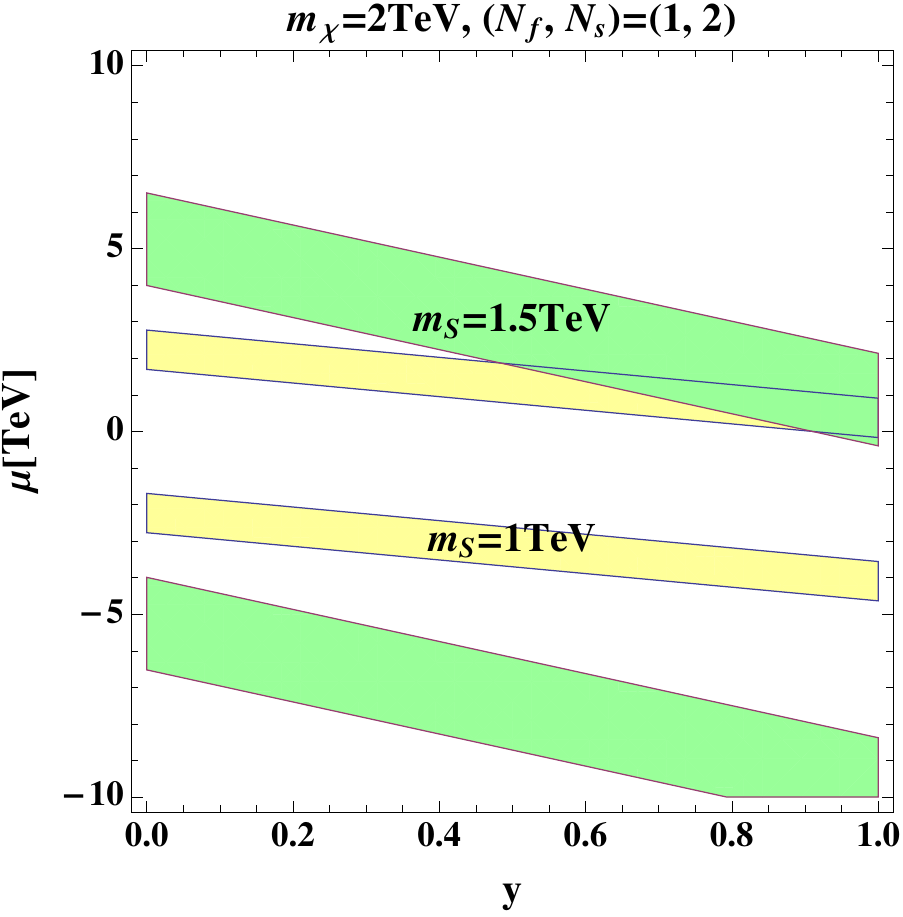}
\end{center}
\caption{Same as Fig. \ref{rhoy}, but for the triplet fermion case.}
\label{chiy}
\end{figure}

Now we turn to the triplet fermion case. The crucial difference from the singlet case is that now the charged members of $\chi$ also contribute to the diphoton channel. From eq.~(\ref{DWchi}), one can roughly estimate that $\BR(\Phi\to\gamma\gamma)$ is a constant, since
\begin{equation}
\frac{\Gamma_{\Phi\to\gamma\gamma}^\chi}{\Gamma_{\Phi\to gg}^\chi}=
\frac{(N_c^2-1)^2\alpha^2}{32N_c^2\alpha_s^2}=\frac{2\alpha^2}{9\alpha_s^2}\approx10^{-3}.
\end{equation}
Therefore, to explain the diphoton excess with $\sigma_{\gamma\gamma}\sim3-10~\fb$, we just need $\sigma_{gg}\sim 3-10~\pb$. The cross sections $\sigma_{\gamma\gamma,gg}$ versus the trilinear coupling $\mu$ are presented in Fig.~\ref{chimu}, while Fig.~\ref{chiy} shows the allowed parameter space in the $y-\mu$ plane. We summarize the properties as follows:
\begin{itemize}
\item The diphoton excess can also be explained naturally with $\mathcal{O}(0.1)$ Yukawa coupling and $\mathcal{O}(\TeV)$ trilinear coupling $\mu$ in the case of triplet fermions $\chi$.
\item The lower and upper bound on $\mu$ comes respectively from $\sigma_{\gamma\gamma}$ and $\sigma_{gg}$, since $\BR(\Phi\to\gamma\gamma)$ is a constant in this case. This behavior is different from the singlet $\rho$ case.
\item It is clear that $\mu<0$ must be satisfied with a bigger $y=0.75$ in the case of $(N_f,N_s)=(2,1)$. Meanwhile for the case of $(N_f,N_s)=(1,2)$, $\mu>0$ is still allowed at $y=0.75$.
\item The simple relation between $\Gamma_{\phi\to\gamma\gamma}^\chi$ and $\Gamma_{\phi\to gg}^\chi$ implies that the allowed bands in the $y-\mu$ plane present a linear correlation. This is clearly shown in Fig.~\ref{chiy}. Moreover, a larger $m_S$ leads to an increase of the slope, and for the same $m_S$ the slope for the $(N_f,N_s)=(2,1)$ scenario is steeper than that of $(N_f,N_s)=(1,2)$.
\end{itemize}

%%%%%%%%%%%%%%%%%%%%%%%
\section{conclusion}
%%%%%%%%%%%%%%%%%%%%%%%

We have interpreted the $750~\GeV$ diphoton excess in the framework of color-octet neutrino mass model. In addition to a singlet scalar $\Phi$ that plays the role of the $750~\GeV$ resonance, the model introduces $N_s$ species of color-octet, electroweak doublet scalars $S$ and $N_f$ color-octet fermions that can be an electroweak singlet $\rho$ or triplet $\chi$. The diphoton signal results from the production of $\Phi$ via gluon fusion through color-octet messengers and its subsequent decay into the diphoton through interactions with charged color-octet particles. We find that for the triplet case $\BR(\Phi\to\gamma\gamma)$ is a constant of about $0.1\%$, while for the singlet case $\BR(\Phi\to\gamma\gamma)$ can be varied. With $\mathcal{O}(0.1)$ Yukawa coupling $y$ and $\mathcal{O}(\TeV)$ trilinear coupling $\mu$, both cases can explain the diphoton excess naturally without conflict with the current experimental limits. Meanwhile, the interference effect between the scalar and fermion octet particles plays a crucial role. The distinct features and allowed parameter space are discussed in detail for both cases, as well as for the two minimal scenarios of neutrino mass generation with $(N_f,N_s)=(2,1)$ or $(1,2)$.

%%%%%%%%%%%%%%%%%%%%%%%
\section*{Acknowledgement}
%%%%%%%%%%%%%%%%%%%%%%%
This work was supported in part by the Grants No. NSFC-11025525, No. NSFC-11575089 and by the CAS Center for Excellence in Particle Physics (CCEPP). Part of numerical analysis was done with the HPC Cluster of SKLTP/ITP-CAS.

%%%%%%%%%%%%%%%%%%%%%%%%%%%%%

\end{document}